\documentclass[aps,prd,preprintnumbers,superscriptaddress,nofootinbib,notitlepage]{revtex4-2}
\usepackage[pdftex]{graphicx}
\usepackage{bm,latexsym,amsmath,amssymb,amsfonts,mathrsfs,bbm}
\usepackage{mathtools}
\usepackage{color}
\allowdisplaybreaks[1]
\usepackage[pdftex,colorlinks=true,linkcolor=blue,citecolor=cyan]{hyperref}
\newcommand*{\D}{{\rm d}}

\newcommand*{\dc}{\mathcal{D}}
\newcommand{\tri}{{}^{(3)}\!e}

\newcommand{\tridot}{{}^{(3)}\!\dot{e}}

\begin{document}
\title{Consistency of higher derivative couplings to matter fields in scalar-tensor gravity}
\author{Tact~Ikeda}
\email[Email: ]{tact@rikkyo.ac.jp}
\affiliation{Department of Physics, Rikkyo University, Toshima, Tokyo 171-8501, Japan
}
\author{Kazufumi~Takahashi}
\email[Email: ]{kazufumi.takahashi@yukawa.kyoto-u.ac.jp}
\affiliation{
Center for Gravitational Physics and Quantum Information,
Yukawa Institute for Theoretical Physics, Kyoto University, 606-8502, Kyoto, Japan
}
\author{Tsutomu~Kobayashi}
\email[Email: ]{tsutomu@rikkyo.ac.jp}
\affiliation{Department of Physics, Rikkyo University, Toshima, Tokyo 171-8501, Japan
}
\begin{abstract}
Recently, a generalization of invertible disformal transformations containing higher-order derivatives of a scalar field has been proposed in the context of scalar-tensor theories of gravity.
By applying this generalized disformal transformation to the Horndeski theory, one can obtain the so-called generalized disformal Horndeski (GDH) theories which are more general healthy scalar-tensor theories than ever.
However, it is unclear whether or not the
GDH theories can be coupled consistently to matter fields because introducing matter fields could break the degeneracy conditions of higher-order scalar-tensor theories and hence yield the unwanted Ostrogradsky ghost.
We investigate this issue and explore the conditions under which a minimal coupling to a matter field is consistent in the
GDH theories without relying on any particular gauge such as the unitary gauge.
We find that all the higher derivative terms in the generalized disformal transformation are prohibited to avoid the appearance of an extra degree of freedom in a generic gauge.
Our analysis shows that, if one considers matter-coupled GDH theories, 
an extra degree of freedom shows up, though it might be a harmless nonpropagating mode when the scalar field has a timelike gradient.
\end{abstract}
\preprint{RUP-23-2, YITP-23-11}
\maketitle
\section{Introduction}

Extending general relativity allows us to study various unresolved issues in the Universe. For instance,
the mechanism of the accelerated expansion of the late Universe is yet unknown,
which motivates the active study of modified gravity as an alternative to dark energy. A modification of gravity at high energies is also strongly motivated
because general relativity is considered a low-energy effective theory. It is thus interesting to explore physics beyond general relativity in a strong gravity regime.
Furthermore, modified gravity models are useful also for comparison with general relativity in the context of testing gravity (see, e.g., Refs.~\cite{Koyama:2015vza,Ferreira:2019xrr,Arai:2022ilw} for reviews).
For instance, in recent years, the direct detection of gravitational waves~\cite{LIGOScientific:2016aoc} and imaging of black hole shadows~\cite{EventHorizonTelescope:2019dse} have been successfully achieved, making it increasingly feasible to test gravity in the strong-field regime.

Given a huge variety of modified theories of gravity to tackle different problems in gravitational and cosmological physics and to be tested against observations and experiments, it is practically impossible to examine each theory individually. It is therefore highly desirable to construct as general a framework as possible to handle many different theories of gravity in a unifying manner.
We are thus motivated to consider a general framework for modified theories of gravity and then limit the theory space based on some criteria that variable theories must satisfy.

Modified gravity is described, at least effectively, by theories equipped with new gravitational degree(s) of freedom (DOFs) on top of the massless graviton DOFs, and as such scalar-tensor theories are often studied where a scalar DOF is taken into account.
Once one goes beyond general relativity,
one may naturally consider higher derivative terms in the gravitational Lagrangian,
but higher-derivative theories are plagued by the Ostrogradsky instability in general~\cite{Woodard:2015zca,Motohashi:2020psc,Aoki:2020gfv}.
Therefore, in generalizing theories of gravity, one must be careful not to induce such an instability arising from higher derivatives.
Even if the Lagrangian itself contains higher derivatives, this instability can be avoided as long as the field equations are intrinsically of second order.
The Horndeski theory~\cite{Horndeski:1974wa,Deffayet:2011gz,Kobayashi:2011nu} is the most general scalar-tensor theory having second-order Euler-Lagrange equations, and due to this merit it has been studied extensively over the recent years.
However, the Horndeski theory is not the most general scalar-tensor theory that is free from the Ostrogradsky instability. The point here is that, if the system is degenerate, higher-order field equations contain less number of DOFs than anticipated from the order of derivatives~\cite{Motohashi:2014opa,Motohashi:2016ftl,Klein:2016aiq,Motohashi:2017eya,Motohashi:2018pxg}.
Degenerate higher-order scalar-tensor (DHOST) theories~\cite{Langlois:2015cwa,Crisostomi:2016czh,BenAchour:2016fzp} exploit this loophole and extend the Horndeski theory to a large family of higher-derivative scalar-tensor theories having a single scalar and two tensor DOFs (see Refs.~\cite{Langlois:2018dxi,Kobayashi:2019hrl} for reviews).

An invertible transformation is a useful tool to construct such a general framework of gravitational theories. Actually,
given that two theories related via invertible field redefinition have the same number of dynamical DOFs~\cite{Domenech:2015tca,Takahashi:2017zgr},
an invertible metric redefinition is a convenient and useful way of generating nontrivial class of healthy scalar-tensor theories from existing ones that are manifestly ghost-free. In particular, derivative-dependent transformations yield higher-derivative theories that are nevertheless free from the Ostrogradsky ghost.
For example, by applying to the Horndeski theory a disformal transformation, which is a general metric redefinition involving the first derivative of the scalar DOF~\cite{Bekenstein:1992pj}, we can obtain a certain subset of DHOST theories~\cite{BenAchour:2016cay}.
Indeed, the first example of scalar-tensor theories beyond Horndeski was obtained in that way~\cite{Zumalacarregui:2013pma}.
It is important to note that, among subclasses of DHOST theories that are systematically constructed by imposing the degeneracy conditions,
only the subclass generated via invertible disformal transformation from the Horndeski theory is physically interesting because
cosmological solutions can be stable (and tensor perturbations remain dynamical)
only in that subclass~\cite{Langlois:2017mxy,deRham:2016wji}.\footnote{A noninvertible subclass of disformal transformations can also be used to generate a certain subclass of DHOST theories, but this subclass does not accommodate stable cosmological solutions or otherwise the tensor perturbations are nondynamical~\cite{Takahashi:2017pje,Langlois:2018jdg}.}
Such disformally generated DHOST theories include, e.g., the so-called ``beyond Horndeski'' or
GLPV theories~\cite{Gleyzes:2014dya,Gleyzes:2014qga} as specific cases.

Recently, a generalization of invertible disformal transformations involving second (and higher) derivatives of the scalar DOF was proposed~\cite{Takahashi:2021ttd}.
By applying this novel class of generalized disformal transformations to the Horndeski theory,
we can derive yet more general higher-order scalar-tensor theories than ever constructed,
which we call generalized disformal Horndeski (GDH) theories~\cite{Takahashi:2022mew}.
Since two theories thus related via invertible disformal transformation are equivalent,
GDH theories are also free from the Ostrogradsky instability (even though the field equations are apparently of higher order).
However, this statement should be taken with care.
The equivalence holds only in vacuum,
and the Horndeski theory with minimally coupled matter and GDH theories with minimally coupled matter are \textit{not} equivalent.
Therefore, the inclusion of (minimally coupled) matter fields in GDH theories could result in extra ghost DOFs.
The possible appearance of extra DOFs can be understood by
moving back to the ``Horndeski frame'' where the gravity sector is described by the Horndeski theory and
the matter sector is coupled to a generalized disformal metric involving second derivatives of the scalar DOF.
In the Horndeski frame, the matter fields are thus coupled with second derivatives of the scalar DOF, 
possibly breaking the degeneracy conditions.
This issue has been pointed out already in the context of DHOST theories~\cite{Deffayet:2020ypa,Garcia-Saenz:2021acj}.
Motivated by this concern,
the consistency of matter couplings in GDH theories has been investigated in Refs.~\cite{Takahashi:2022mew,Naruko:2022vuh,Takahashi:2022ctx}.
In those previous studies, however,
the unitary gauge is taken, which is justified only if the scalar field has a timelike gradient.
The conditions that remove extra ghost DOFs derived in Refs.~\cite{Takahashi:2022mew,Naruko:2022vuh,Takahashi:2022ctx} should therefore be weaker than those that would be derived without assuming any particular gauge.
In other words, upon imposing the degeneracy conditions validated only in the unitary gauge,
there would be an apparent Ostrogradsky ghost away from the unitary gauge.
Such a gauge-dependent Ostrogradsky ghost would be a nonpropagating ``shadowy mode,'' and hence harmless~\cite{DeFelice:2018ewo,DeFelice:2021hps}.
Having said that, we need to remove such an extra DOF in any gauge to safely consider, e.g., static stars and black holes dressed with a static scalar profile.
Also, since such theories are rather tricky and the shadowy mode requires a careful treatment,
we are interested mostly in theories without the shadowy mode.
Along this line of thought, in this paper, we explore, without assuming any particular gauge, the conditions under which matter couplings to GDH theories are consistent and there is no extra DOF.

This paper is organized as follows.
In the next section, we briefly review the generalized disformal transformation~\cite{Takahashi:2021ttd} and
introduce the action for GDH theories by applying a generalized disformal transformation to the Horndeski theory.
In Sec.~\ref{sec:Consistency of matter coupling}, we study the consistency of matter couplings in the GDH theory.
We review the previous results obtained in the unitary gauge~\cite{Takahashi:2022mew,Naruko:2022vuh,Takahashi:2022ctx}, and then investigate the conditions for the matter couplings to be consistent away from the unitary gauge.
We finally draw our conclusions in Sec.~\ref{sec:Conclusions}.

\section{Generalized disformal transformations}

\subsection{Invertible disformal transformations with higher-order derivatives}

Let us consider a generalized disformal transformation defined as~\cite{Takahashi:2021ttd}
\begin{align}
	\bar g_{\mu\nu}=F_0 g_{\mu\nu}+F_1 \phi_\mu\phi_\nu+2F_2\phi_{(\mu}X_{\nu)}+F_3X_{\mu}X_{\nu},\label{eq:GeneralizedDisformalTransformation}
\end{align}
where $\phi_\mu\coloneqq \nabla_\mu\phi$ and $X_\mu\coloneqq \nabla_\mu X$ with $X\coloneqq \phi^\mu\phi_\mu$.
Here, $F_i$ ($i=0,1,2,3$) are functions of $(\phi,X,Y,Z)$, where $Y\coloneqq \phi^\mu X_\mu$ and $Z\coloneqq X^\mu X_\mu$.
Restricting the form of the functions to be $F_0=F_0(\phi,X)$, $F_1=F_1(\phi,X)$ and $F_2=F_3=0$, we obtain the conventional disformal transformation~\cite{Bekenstein:1992pj}
\begin{align}
	\bar g_{\mu\nu}=F_0(\phi,X) g_{\mu\nu}+F_1(\phi,X) \phi_\mu\phi_\nu.\label{eq:ConventionalDisformalTransformation}
\end{align}
In this sense, the transformation~\eqref{eq:GeneralizedDisformalTransformation} involves the conventional one~\eqref{eq:ConventionalDisformalTransformation}.

Following Ref.~\cite{Takahashi:2021ttd}, we summarize the conditions under which the transformation~\eqref{eq:GeneralizedDisformalTransformation} is invertible.
The essential ingredient of the invertible generalized disformal transformation is
the requirement that a set of generalized disformal transformations forms a group under the following two operations:
\begin{align}
	\left(\bar{g} \cdot \hat{g}\right)_{\mu \nu}\coloneqq \bar{g}_{\mu \alpha} g^{\alpha \beta} \hat{g}_{\beta \nu},
    \quad \text{(Matrix product)}
\end{align}
and
\begin{align}
	\left(\bar{g} \circ \hat{g}\right)_{\mu \nu}[g, \phi]\coloneqq \bar{g}_{\mu\nu}[\hat{g},\phi].
    \quad \text{(Functional composition)}
\end{align}
We are then allowed to construct the inverse of the transformed metric and the inverse transformation.
Note that, in contrast to the case of conventional disformal transformation~\eqref{eq:ConventionalDisformalTransformation}, the closedness under the functional composition for the generalized disformal transformation~\eqref{eq:GeneralizedDisformalTransformation} is nontrivial as the transformation law involves the derivative of the metric.
For the set of generalized disformal transformations to form a group, it is sufficient that the following conditions are satisfied~\cite{Takahashi:2021ttd}:
\begin{align}
	F_{0} \neq 0, \quad \mathcal{F} \neq 0, \quad \bar{X}_{X} \neq 0, \quad \bar{X}_{Y}=\bar{X}_{Z}=0, \quad\left|\frac{\partial(\bar{Y}, \bar{Z})}{\partial(Y, Z)}\right| \neq 0,\label{eq:InvertibilityConditions}
\end{align}
where
\begin{align}
    \bar{X}\coloneqq \bar{g}^{\mu \nu} \phi_{\mu} \phi_{\nu},
    \quad
    \bar{Y}\coloneqq \bar{g}^{\mu \nu} \phi_{\mu} \bar{X}_{\nu},
    \quad
    \bar{Z}\coloneqq \bar{g}^{\mu \nu} \bar{X}_{\mu} \bar{X}_{\nu},
\end{align}
and 
\begin{align}
    \mathcal{F}(\phi,X,Y,Z)
    &\coloneqq F_0^2+F_0(XF_1+2YF_2+ZF_3)+(Y^2-XZ)(F_2^2-F_1F_3).\label{eq:CalF}
\end{align}
Note that, among the conditions in Eq.~\eqref{eq:InvertibilityConditions}, the one that guarantees the closedness under the functional composition is $\bar{X}_{Y}=\bar{X}_{Z}=0$.
Suppose that these conditions are satisfied.
The inverse metric~$\bar g^{\mu\nu}$ of $\bar g_{\mu\nu}$ is then given by
\begin{align}
	\bar g^{\mu\nu}=f_0 g^{\mu\nu}+f_1 \phi^\mu\phi^\nu+2f_2\phi^{(\mu}X^{\nu)}+f_3X^{\mu}X^{\nu},\label{eq:InverseMetric}
\end{align}
where
\begin{align}
    f_0&\coloneqq \frac{1}{F_{0}},
        \quad
        f_1\coloneqq  -\frac{F_{0} F_{1}-Z\left(F_{2}^{2}-F_{1} F_{3}\right)}{F_{0} \mathcal{F}}, \quad 
        f_2\coloneqq -\frac{F_{0} F_{2}+Y\left(F_{2}^{2}-F_{1} F_{3}\right)}{F_{0} \mathcal{F}},
        \quad
        f_3\coloneqq -\frac{F_{0} F_{3}-X\left(F_{2}^{2}-F_{1} F_{3}\right)}{F_{0} \mathcal{F}}.
        \label{eq:CoefficientOfInverseMetric}
\end{align}
The following formula is also useful for reconstructing the barred metric~$\bar g_{\mu\nu}$ [Eq.~\eqref{eq:GeneralizedDisformalTransformation}]
when its inverse is given in the form of Eq.~\eqref{eq:InverseMetric}:
\begin{align}
    F_0&\coloneqq \frac{1}{f_{0}}, \quad
    F_1\coloneqq  -\frac{f_{0} f_{1}-Z\left(f_{2}^{2}-f_{1} f_{3}\right)}{f_{0} \mathcal{H}}, \quad 
    F_2\coloneqq -\frac{f_{0} f_{2}+Y\left(f_{2}^{2}-f_{1} f_{3}\right)}{f_{0} \mathcal{H}}, \quad
    F_3\coloneqq -\frac{f_{0} f_{3}-X\left(f_{2}^{2}-f_{1} f_{3}\right)}{f_{0} \mathcal{H}}.
    \label{eq:ReconstructBarredMetric}
\end{align}
Here, we have defined
\begin{align}
    \mathcal{H}(\phi,X,Y,Z)
    &\coloneqq f_0^2+f_0(Xf_1+2Yf_2+Zf_3)+(Y^2-XZ)(f_2^2-f_1f_3),
    \label{calH_f}
\end{align}
which is obtained simply by replacing $F_i$ in Eq.~\eqref{eq:CalF} by $f_i$.
Having constructed the inverse metric explicitly, let us next look at
the inverse transformation.
The inverse transformation of~\eqref{eq:GeneralizedDisformalTransformation} is expressed as
\begin{align}
	g_{\mu \nu}[\bar g,\phi]&= \frac{1}{F_{0}} \bar{g}_{\mu \nu}-\frac{\bar{X}_{X}^{2} F_{1}-2 \bar{X}_{\phi} \bar{X}_{X} F_{2}+\bar{X}_{\phi}^{2} F_{3}}{\bar{X}_{X}^{2} F_{0}} \phi_{\mu} \phi_{\nu}
	-2 \frac{\bar{X}_{X} F_{2}-\bar{X}_{\phi} F_{3}}{\bar{X}_{X}^{2} F_{0}} \phi_{(\mu} \bar{X}_{\nu)}-\frac{F_{3}}{\bar{X}_{X}^{2} F_{0}} \bar{X}_{\mu} \bar{X}_{\nu},\label{eq:InverseTransformation}
\end{align}
where $F_i$'s in the right-hand side are given as functions of $(\phi,\bar X,\bar Y,\bar Z)$.
Thanks to the group structure of the set of generalized disformal transformations, one can thus obtain the inverse metric and inverse transformation.

\subsection{Generalized disformal Horndeski theories}\label{sec:Generalized disformal Horndeski theories}

We now define a scalar-tensor theory obtained by applying a generalized disformal transformation to the metric in the Horndeski theory.
The Horndeski theory (in vacuum) is described by the action~\cite{Horndeski:1974wa,Deffayet:2011gz,Kobayashi:2011nu}
\begin{align}
    S_{\mathrm{Hor}}[g_{\mu\nu},\phi]\coloneqq \int \D^4x \sqrt{-g}\,\mathcal{L}_{\mathrm{Hor}}[g_{\mu\nu},\phi],
\end{align}
where
\begin{align}
    \mathcal{L}_{\mathrm{Hor}}&\coloneqq 
    G_2(\phi,X)+G_3(\phi,X)\Box \phi+
    G_4(\phi,X)R-2G_{4,X}(\phi,X)[(\Box\phi)^2-\phi^{\mu\nu}\phi_{\mu\nu}]
    \notag \\ &\quad 
    +G_5(\phi,X)G^{\mu\nu}\phi_{\mu\nu}+\frac{1}{3}G_{5,X}(\phi,X)[(\Box\phi)^3-3\Box\phi\phi^{\mu\nu}\phi_{\mu\nu}+2\phi_{\mu\nu}\phi^{\mu\rho}\phi^\nu_\rho],
\end{align}
with $\phi_{\mu\nu}\coloneqq \nabla_\mu\nabla_\nu\phi$.
Performing a generalized disformal transformation satisfying the invertibility conditions~\eqref{eq:InvertibilityConditions}, we obtain a new action for a scalar-tensor theory~\cite{Takahashi:2022mew}
\begin{align}
    S_{\mathrm{GDH}}[g_{\mu\nu},\phi]\coloneqq S_{\mathrm{Hor}}[\bar g_{\mu\nu},\phi]=\int \D^4x \sqrt{-g}
    \,\mathcal{J}
    \mathcal{L}_{\mathrm{Hor}}[\bar g_{\mu\nu}, \phi],
\end{align}
where we have defined
\begin{align}
    \mathcal{J}\coloneqq \frac{\sqrt{-\bar g}}{\sqrt{-g}}=F_0\mathcal{F}^{1/2}=f_0^{-1}\mathcal{H}^{-1/2}.
\end{align}
This theory was dubbed the generalized disformal Horndeski (GDH) theory~\cite{Takahashi:2022mew}.
Since the generalized disformal transformation is just a field redefinition,
$S_{\mathrm{GDH}}[g_{\mu\nu},\phi]$ and $S_{\mathrm{Hor}}[\bar g_{\mu\nu},\phi]$ are mathematically equivalent as long as the transformation satisfies the invertibility conditions~\eqref{eq:InvertibilityConditions}.
This in particular means that there are one scalar and two tensor DOFs in the (vacuum) GDH theory as in the (vacuum) Horndeski theory, even though the field equations in the former theory contain higher derivatives in general.
However, in the presence of matter fields, things become subtle and the relation between the two theories must be examined carefully.

To see this point more closely, let us add matter field(s) (collectively denoted by $\Psi$) minimally coupled to the GDH theory,
\begin{align}
    S_{\mathrm{GDH}}[g_{\mu\nu},\phi]+S_{\mathrm{m}}[ g_{\mu\nu},\Psi]=S_{\mathrm{Hor}}[\bar g_{\mu\nu},\phi]+S_{\mathrm{m}}[g_{\mu\nu},\Psi].\label{action-with-matter}
\end{align}
We see that the matter fields are coupled with the generalized disformal metric~\eqref{eq:InverseTransformation} in the Horndeski frame (the right-hand side) in which the gravitational part of the action is written manifestly in the Horndeski form in terms of the metric~$\bar g_{\mu\nu}$.
Since the generalized disformal metric~\eqref{eq:InverseTransformation} contains higher-order derivatives of $\phi$, there is no guarantee that the coupling to matter is consistent, i.e., no unwanted DOF appears as an Ostrogradsky ghost through this coupling.
If such an additional dangerous DOF were to appear, then the GDH theory in the presence of (minimally coupled) matter would be inconsistent, albeit healthy in vacuum.\footnote{We expect that the mass of the Ostrogradsky mode would be proportional to some negative power of the energy density of the matter field.
Therefore, from the EFT point of view, the ghost would be irrelevant if we push its mass above the cutoff scale.}
In the next section, we will study this point in detail.

\section{Consistency of matter coupling}\label{sec:Consistency of matter coupling}

\subsection{Unitary gauge}

In this subsection, we briefly review the consistency of matter coupling in the GDH theory under the unitary gauge, where the scalar field is spatially uniform.
Though it is not always justified, one is allowed to take the unitary gauge at least in the context of cosmology where the scalar field is supposed to have a timelike gradient.
The case of bosonic matter fields was discussed in Refs.~\cite{Takahashi:2022mew,Naruko:2022vuh,Takahashi:2022ctx}.
As we saw in Sec.~\ref{sec:Generalized disformal Horndeski theories}, in the Horndeski frame, the matter fields are coupled to the generalized disformal metric (with respect to the metric that describes the gravity sector), and hence the matter action involves higher-order derivatives of $\phi$.
This indicates that the matter action yields the time derivative of the lapse function under the unitary gauge, which generically makes the (otherwise nondynamical) lapse function dynamical.
Thus, the GDH theory would give rise to the Ostrogradsky ghost in general in the presence of matter fields.
Fortunately, one can remove the Ostrogradsky ghost by restricting the generalized disformal metric to the following form~\cite{Takahashi:2022mew,Naruko:2022vuh,Takahashi:2022ctx}:
\begin{align}
    \bar{g}_{\mu\nu} = \tilde{F}_0g_{\mu\nu} +\tilde{F}_1\phi_\mu\phi_\nu +2\tilde{F}_2\phi_{(\mu}\mathcal{X}_{\nu)} +\tilde{F}_3{\cal X}_{\mu}\mathcal{X}_{\nu},
    \qquad
    \mathcal{X}_{\mu}\coloneqq \left(\delta_\mu^\alpha-\frac{\phi_\mu\phi^\alpha}{X}\right)\partial_\alpha X,
    \label{GDT_consistent_unitary}
\end{align}
where ${\cal X}_\mu$ is the derivative of $X$ projected onto a constant-$\phi$ hypersurface and $\tilde{F}_i=\tilde{F}_i(\phi,X,{\cal Z})$, with
\begin{align}
    {\cal Z}\coloneqq {\cal X}^\mu{\cal X}_\mu
    =Z-\frac{Y^2}{X}.
\end{align}
The point is that the object~${\cal X}_\mu$ does not contain the time derivative of the lapse function under the unitary gauge where $\phi=\phi(t)$.
Note that Eq.~\eqref{GDT_consistent_unitary} is embedded in the original generalized disformal transformation~\eqref{eq:GeneralizedDisformalTransformation} as
    \begin{align}
    F_0=\tilde{F}_0, \qquad
    F_1=\tilde{F}_1-\frac{2Y}{X}\tilde{F}_2+\frac{Y^2}{X^2}\tilde{F}_3, \qquad
    F_2=\tilde{F}_2-\frac{Y}{X}\tilde{F}_3, \qquad
    F_3=\tilde{F}_3.
    \end{align}
On the other hand, the case of fermionic matter fields needs a separate treatment as the matter action is written in terms of the tetrad rather than the metric itself.
The authors of Ref.~\cite{Takahashi:2022ctx} developed the transformation law for the tetrad to show that the consistency of fermionic matter coupling requires an additional condition~$\tilde{F}_3=0$~\cite{Takahashi:2022ctx}.

Here, it should be emphasized that all these results were obtained in the unitary gauge, and hence the scalar field was assumed to have a timelike gradient.
Away from the unitary gauge, apparently there would be an Ostrogradsky mode, but this mode could be harmless because
it would be a nonpropagating ``shadowy mode'' that satisfies a three-dimensional elliptic differential equation on a spacelike hypersurface, leading to a configuration completely determined by boundary conditions.
(See Refs.~\cite{DeFelice:2018ewo,DeFelice:2021hps,DeFelice:2022xvq} for a more detailed discussion in the context of U-DHOST theories.)
Having said that, by removing the Ostrogradsky mode in any gauge, one may safely consider, for example, static stars and black holes dressed with a static scalar field.
Also, we need a careful treatment for the shadowy mode, and hence theories without the shadowy mode are still of primary interest to us.
In what follows, we investigate the consistency of matter coupling away from the unitary gauge and derive degeneracy conditions without any extra mode.

\subsection{Away from the unitary gauge}\label{ssec:AwayFromUnitary}
Let us now explore the consistency of the matter coupling away from the unitary gauge.
We start with the Horndeski-frame Lagrangian
\begin{align}
    \mathcal{L}[g_{\mu\nu},\phi]=\mathcal{L}_{\mathrm{Hor}}[g_{\mu\nu},\phi]+\mathcal{L}_{\mathrm{m}}[\bar g_{\mu\nu},\psi],
    \label{eq:Horndeski-frame_Lagrangian}
\end{align}
where $\bar g_{\mu\nu}$ is defined in Eq.~\eqref{eq:GeneralizedDisformalTransformation}. Note that we have interchanged the roles of $g_{\mu\nu}$ and $\bar g_{\mu\nu}$ as compared to those in Eq.~\eqref{action-with-matter}.
In any case, so long as the (generalized) disformal transformation is invertible, the barred metric is some disformal transformation of the unbarred metric and vice versa, and hence this is just a matter of convention.
Note, however, that the Lagrangian~\eqref{action-with-matter} makes sense
as a theory of gravity nonminimally coupled to matter even if the disformal transformation is noninvertible, though one cannot move to the equivalent description in the GDH (or Jordan) frame in this case.
Therefore, in this subsection, we study a system described by the Lagrangian~\eqref{action-with-matter} without imposing the invertibility conditions for the disformal transformation from the outset.\footnote{Precisely speaking, we assume the first two conditions in Eq.~\eqref{eq:InvertibilityConditions} to guarantee the existence of the barred inverse metric~$\bar{g}^{\mu\nu}$, but do not necessarily impose the last three conditions. The authors of Ref.~\cite{Naruko:2022vuh} used a similar approach to specify the degeneracy conditions under the unitary gauge without imposing the invertibility conditions from the outset.}
For simplicity, we assume that the matter sector is described by a massless scalar field~$\psi$,
\begin{align}
    S_{\mathrm{m}}[\bar g_{\mu\nu},\psi]
    &\coloneqq -\frac{1}{2}\int \D^4x\sqrt{-\bar g}\,\bar g^{\mu\nu}\psi_\mu \psi_\nu
    =-\frac{1}{2}\int \D^4x\sqrt{-g}\,\mathcal{J}\bar g^{\mu\nu}\psi_\mu \psi_\nu,
\end{align}
where $\psi_\mu\coloneqq \nabla_\mu \psi$ and $\bar g^{\mu\nu}$ is the inverse metric associated with $\bar g_{\mu\nu}$ defined in Eq.~\eqref{eq:InverseMetric}.

We expect that the system described by the Lagrangian~\eqref{action-with-matter} has four physical DOFs, where three come from the gravity sector ($g_{\mu\nu}$ and $\phi$) and one from the matter scalar field~$\psi$.
In order to avoid an unwanted fifth DOF, terms with the highest time derivatives must possess a degenerate structure.
In order to study the kinetic structure of the Lagrangian~\eqref{action-with-matter} in detail, let us introduce the Arnowitt-Deser-Misner (ADM) variables as
    \begin{align}
    g_{\mu\nu} \D x^\mu \D x^\nu=-N^2\D t^2+h_{ij}(\D x^i+N^i\D t)(\D x^j+N^j\D t), \label{ADM}
    \end{align}
where $N$ is the lapse function, $N^i$ is the shift vector, and $h_{ij}$ is the induced metric.
Note that we do not choose the unitary gauge, and hence the timelike unit normal vector~$n_\mu=-N\delta^0_\mu$ associated with a constant-$t$ hypersurface is not proportional to $\phi_\mu$.
The extrinsic curvature is written in terms of the ADM variables as
    \begin{align}
    K_{ij}=\frac{1}{2N}\left(\dot{h}_{ij}-{\rm D}_iN_j-{\rm D}_jN_i\right),
    \end{align}
where a dot denotes the time derivative and ${\rm D}_i$ denotes the covariant derivative associated with $h_{ij}$.
We also define the variables associated with the first and second time derivatives of $\phi$ and the first time derivative of $\psi$ as follows:
    \begin{align}
    A_*\coloneqq n^\mu \nabla_\mu \phi, \qquad
    X_*\coloneqq n^\mu \nabla_\mu X, \qquad
    \psi_*\coloneqq n^\mu \nabla_\mu \psi.\label{eq:Astar}
    \end{align}
The kinetic structure of the Lagrangian~\eqref{action-with-matter} can be captured by the Hessian matrix~$\mathbb{H}$ of the Lagrangian~\eqref{eq:Horndeski-frame_Lagrangian} with respect to $K_{ij}$, $X_*$, and $\psi_*$.
Written explicitly, one has
\begin{align}
	\mathbb{H}=
	\begin{pmatrix}
		\mathcal{K}^{i j, k l} & 0 & 0
		\\
		0 & \mathcal{A} & \mathcal{M}
		\\
		0 & \mathcal{M} & \mathcal{P}
	\end{pmatrix}
	,\label{eq:Hessian}
\end{align}
with
\begin{align}
    \mathcal{K}^{ij,kl}\coloneqq \frac{\partial^2\mathcal{L}}{\partial K_{ij}\partial K_{kl}}, \qquad
    \mathcal{A}\coloneqq \frac{\partial^2\mathcal{L}}{\partial X_*^2}, \qquad
    \mathcal{M}\coloneqq \frac{\partial^2\mathcal{L}}{\partial X_*\partial \psi_*}, \qquad
    \mathcal{P}\coloneqq \frac{\partial^2\mathcal{L}}{\partial \psi_*^2}.
    \label{eq:Hessian_components}
\end{align}
Note that there is no kinetic mixing between the gravitational and matter sectors:
The gravitational (Horndeski) sector concerns only $\mathcal{K}^{ij,kl}$, while the matter sector concerns only $\mathcal{A}$, $\mathcal{M}$, and $\mathcal{P}$.
In order to kill the unwanted DOF revived due to the matter coupling, we require that the $2\times 2$ lower-right submatrix of $\mathbb{H}$ is degenerate, i.e.,
\begin{align}
	\dc\coloneqq \mathcal{A}\mathcal{P}-\mathcal{M}^2=0.
\end{align}
The quantity~$\dc$ can be rewritten in the form of a polynomial in $\{A_*,X_*,\psi_*,\mathcal{Q}_1,\mathcal{Q}_2,\mathcal{Q}_3\}$ as
\begin{align}
    \dc=\sum_{i,j,k,l,m,n\ge 0} d_{ijklmn}(\phi,X,Y,Z)
    \,A_*^{i}
    \,X_*^{j}
    \,\psi_*^{k}
    \,\mathcal{Q}_1^{l}
    \,\mathcal{Q}_2^{m}
    \,\mathcal{Q}_3^{n},\label{eq:d_ijklmn}
\end{align}
with
\begin{align}
\mathcal{Q}_1\coloneqq g^{\mu\nu}\psi_{\mu}\psi_{\nu},
\quad
\mathcal{Q}_2\coloneqq g^{\mu\nu}\psi_{\mu}\phi_{\nu},
\quad
\mathcal{Q}_3\coloneqq g^{\mu\nu}\psi_{\mu}X_{\nu}.\label{eq:Q123}
\end{align}
Note that the coefficients~$d_{ijklmn}$ depend only on the functions~$f_i$ characterizing the generalized disformal transformation.
In order for $\dc$ to vanish for any configuration of $\phi$ and $\psi$, we shall fix $f_i$'s so that all $d_{ijklmn}$'s vanish.\footnote{Under the unitary gauge, we have $A_*=(-X)^{1/2}$, $X_*=-Y(-X)^{-1/2}$, and $\psi_*=-(-X)^{-1/2}\mathcal{Q}_2$.
Hence, we obtain a weaker condition that the coefficients in front of $\mathcal{Q}_1^{l}\mathcal{Q}_2^{p}\mathcal{Q}_3^{n}$ vanish, i.e., $\sum_{i,j,k\ge 0}(-1)^{j+k}(-X)^{(i-j-k)/2}\,Y^j\,\delta_{k+m,p}\,d_{ijklmn}(\phi,X,Y,Z)=0$ for all $l,n,p\,(\ge 0)$.}
Since the full expression of $\dc$ is extremely involved, we proceed step by step:
Among nonvanishing $d_{ijklmn}$, we first focus on the simplest one(s) to read off condition(s) that $f_i$'s should satisfy.
We then substitute the condition(s) back into $\dc$, which simplifies some of $d_{ijklmn}$'s.
With simplified $d_{ijklmn}$'s, we follow the same steps, selecting the simplest one(s) to find additional condition(s) on $f_i$'s.
Repeating this procedure, we finally obtain a set of conditions on $f_i$'s under which all $d_{ijklmn}$'s vanish, i.e., $\dc=0$.
In what follows, we apply this strategy to fix the functional form of $f_i$'s.
It should be noted that we assume $F_0\ne 0$ and $\mathcal{J}\ne 0$ throughout the following discussion because otherwise one cannot define the barred inverse metric~$\bar{g}^{\mu\nu}$.

The condition that the coefficient of $\psi_*^2$ vanishes
yields
\begin{align}
    f_3=0.
\end{align}
Likewise, from the coefficients of $A_*^4\mathcal{Q}_3^2$ and $X_*^4\mathcal{Q}_2^2$, we find that $f_2$ must be of the form
\begin{align}
	\mathcal{J}f_2=\alpha_2(\phi,X),
\end{align}
where $\alpha_2$ is a function of $\phi$ and $X$ which is arbitrary at this step.
From the coefficients of $\mathcal{Q}_1$ and $A_*^2\mathcal{Q}_1$, we see that $f_0$ must take the form
\begin{align}
	\mathcal{J}f_0=\alpha_0(\phi,X) +\beta_0(\phi,X) Y,
\end{align}
where $\alpha_0$ and $\beta_0$ are arbitrary functions of $\phi$ and $X$.
Then, from the coefficient of $A_*^2\psi_*^2$, we find that
$\alpha_2$ must be related to $\beta_0$ by
\begin{align}
	\alpha_2=-\beta_0.
\end{align}
The coefficient of $A_*X_*^3\mathcal{Q}_2^2$ leads us to the following relation:
\begin{align}
	\beta_0(\mathcal{J}f_1)_{,ZZ}=0.\label{eq:d130020=0}
\end{align}
We now have the two branches of solutions,
$\beta_0=0$ and $(\mathcal{J}f_1)_{,ZZ}=0$.

Let us first consider the branch $\beta_0=0$.
In this case, from the coefficients of $\mathcal{Q}_2^2$, $A_*^2\mathcal{Q}_2^2$, and $A_*^4\mathcal{Q}_2^2$,
we see that $f_1$ must take the form
\begin{align}
	\mathcal{J}f_1=\alpha_1(\phi,X),
\end{align}
where $\alpha_1$ is an arbitrary function of $\phi$ and $X$.
Combining the conditions obtained so far, we obtain the following relations:
\begin{align}
	f_0=\frac{\alpha_0}{\mathcal{J}}
	,\qquad
	f_1=\frac{\alpha_1}{\mathcal{J}}
	,\qquad
	f_2=f_3=0.
\end{align}
On top of these, we have $\mathcal{J}^{-1}=f_0\mathcal{H}^{1/2}$ with $\mathcal{H}$ defined in Eq.~\eqref{calH_f}, which yields
\begin{align}
    \mathcal{J}=\sqrt{\alpha_0^3(\alpha_0+\alpha_1X)}.
\end{align}
We now know $f_i$'s as functions of $(\phi,X)$. Written explicitly, we have
\begin{align}
	f_0=\frac{\alpha_0}{\sqrt{\alpha_0^3(\alpha_0+\alpha_1X)}}
	,\qquad
	f_1=\frac{\alpha_1}{\sqrt{\alpha_0^3(\alpha_0+\alpha_1X)}}
	,\qquad
	f_2=f_3=0.
	\label{eq:consistent_trivial}
\end{align}
By use of Eq.~\eqref{eq:ReconstructBarredMetric}, the barred metric can be reconstructed as
\begin{align}
    \bar{g}_{\mu\nu}=\sqrt{\alpha_0(\alpha_0+\alpha_1X)}\,g_{\mu\nu}-\frac{\alpha_0\alpha_1}{\sqrt{\alpha_0(\alpha_0+\alpha_1X)}}\phi_\mu\phi_\nu. \label{tr-conv}
\end{align}
Note that both coefficients are now functions of $(\phi,X)$, and hence this is nothing but a conventional disformal transformation.

Let us now study the other branch of solutions for Eq.~\eqref{eq:d130020=0}, i.e., $(\mathcal{J}f_1)_{,ZZ}=0$.
After straightforward manipulations, one can show that all $d_{ijklmn}$'s vanish if and only if
    \begin{align}
    f_0=\frac{\alpha_0+\beta_0Y}{\mathcal{J}}, \qquad
    f_1=\frac{\alpha_0\alpha_1+\beta_0^2Z}{\mathcal{J}(\alpha_0+\beta_0Y)}, \qquad
    f_2=-\frac{\beta_0}{\mathcal{J}}, \qquad
    f_3=0, \qquad
    \mathcal{J}^2=\alpha_0(\alpha_0+\alpha_1 X)(\alpha_0+\beta_0Y)^2,
    \label{eq:consistent_noninvertible}
    \end{align}
with $\alpha_0(\ne 0)$, $\alpha_1$, and $\beta_0$ being functions of $(\phi,X)$.
By use of Eq.~\eqref{eq:ReconstructBarredMetric}, the coefficient functions of the barred metric can be reconstructed from Eq.~\eqref{eq:consistent_noninvertible} as
    \begin{align}
    F_0=\sqrt{\alpha_0(\alpha_0+\alpha_1 X)}, \qquad
    F_1=-\frac{\alpha_0\alpha_1}{F_0}, \qquad
    F_2=\frac{\alpha_0\beta_0}{F_0}, \qquad
    F_3=\frac{X\beta_0^2}{F_0},
    \label{eq:consistent_noninvertible_F}
    \end{align}
or written explicitly,\footnote{Our result is consistent with that derived taking the unitary gauge (see Sec.~III of Ref.~\cite{Naruko:2022vuh}).}
\begin{align}
	\bar g_{\mu\nu}=\sqrt{\alpha_0(\alpha_0+\alpha_1 X)}\left[g_{\mu\nu}-\frac{\alpha_1}{\alpha_0+\alpha_1 X}\phi_\mu\phi_\nu+\frac{2\beta_0}{\alpha_0+\alpha_1 X}\phi_{(\mu}X_{\nu)}+\frac{X\beta_0^2}{\alpha_0(\alpha_0+\alpha_1 X)}X_{\mu}X_{\nu}\right].
	\label{eq:consistent_barred_metric}
\end{align}
Interestingly, one can check that the generalized disformal transformation~\eqref{eq:consistent_barred_metric} satisfies the degeneracy condition even in the presence of a k-essence matter scalar field whose Lagrangian is written as a general function of $\psi$ and $\bar g^{\mu\nu}\psi_\mu\psi_\nu$.
Note, however, that the above result does not satisfy a part of the invertibility conditions (specifically, $\bar{X}_Y=\bar{X}_Z=0$) in general.
Indeed, for the above choice of $f_i$'s, we have
    \begin{align}
    \bar{X}=X\left(f_0+Xf_1+2Yf_2+\frac{Y^2}{X}f_3\right)
    =\frac{X\left[\alpha_0^2+X\alpha_0\alpha_1-\beta_0^2(Y^2-XZ)\right]}{\mathcal{J}(\alpha_0+\beta_0Y)},
    \end{align}
which has a nontrivial dependence on $Y$ and $Z$ unless $\beta_0=0$.
If $\beta_0=0$, the generalized disformal transformation~\eqref{eq:consistent_barred_metric} reduces to Eq.~\eqref{tr-conv}, i.e., the conventional one.

So far, we have found that there exists a nontrivial family of generalized disformal metrics described by Eq.~\eqref{eq:consistent_barred_metric} that allows for consistent coupling of a k-essence scalar field without an extra mode.
As mentioned above, this family does not satisfy a part of the invertibility conditions in general, meaning that one cannot move to the equivalent description in the Jordan frame.
The only exception is the case~$\beta_0=0$, where Eq.~\eqref{eq:consistent_barred_metric} reduces to the conventional disformal metric~\eqref{eq:ConventionalDisformalTransformation}.
Nevertheless, even if $\beta_0\ne 0$, the theory makes sense as gravity nonminimally coupled to matter, and hence we could keep it in our consideration.
However, as we shall show in the \hyperref[App]{Appendix}, this family does not allow for consistent coupling of fermionic matter fields unless $\beta_0=0$.
Therefore, all the higher-derivative terms in the generalized disformal metric are prohibited when we require that both bosonic and fermionic matter couplings do not introduce an extra mode, even if we do not impose the invertibility conditions.
Our analysis shows that, if one considers the generalized disformal transformation with nontrivial higher-derivative terms, an extra mode shows up.
When the scalar field has a timelike gradient, this extra mode is nothing but a shadowy mode.
As clarified in Refs.~\cite{DeFelice:2018ewo,DeFelice:2021hps},
the shadowy mode itself is harmless as it satisfies a three-dimensional elliptic differential equation on a spacelike hypersurface and hence does not propagate.
However, this extra mode does propagate around a background with a spacelike gradient of the scalar field, which is problematic.

\section{Conclusions}\label{sec:Conclusions}

In this work,
we have considered a general framework of modified theories of gravity and explored the viable theory space based on the criterion of whether or not gravity can be coupled consistently to matter fields.
To do so,
we have investigated the degeneracy conditions of generalized disformal Horndeski (GDH) theories in the presence of a minimally coupled matter field, which is represented by a canonical scalar field.
We have started with the Horndeski-frame Lagrangian~\eqref{eq:Horndeski-frame_Lagrangian} where the gravitational action is given by the Horndeski one while the matter field is coupled to the generalized disformal metric~\eqref{eq:GeneralizedDisformalTransformation}.
We have rewritten the total Lagrangian in terms of the $3+1$~language and constructed the Hessian matrix so that we can investigate the kinetic structure of the theory.
The degeneracy conditions are required for the matter coupling to be consistent, giving the conditions that the determinant of the Hessian matrix vanishes.
The degeneracy conditions have been written in the form~\eqref{eq:d_ijklmn} as a polynomial in $\{A_*,X_*,\psi_*,\mathcal{Q}_1,\mathcal{Q}_2,\mathcal{Q}_3\}$, which are independent functions of spacetime constructed out of the derivatives of the gravitational scalar field~$\phi$ and the matter scalar field~$\psi$ [see Eqs.~\eqref{eq:Astar} and~\eqref{eq:Q123}].
In order for the degeneracy conditions to be satisfied for arbitrary configurations of $\phi$ and $\psi$, all the coefficients~$d_{ijklmn}$ of the polynomial must vanish.
We have thus arrived at the following conclusions: (i) if one sticks to the invertible transformations satisfying the conditions~\eqref{eq:InvertibilityConditions} so that the equivalent GDH theory in the Jordan frame exists, only the conventional disformal metric can be consistently coupled to the kinetic term of the matter scalar field; (ii) however, if one gives up the invertibility conditions and just considers a scalar field coupled nonminimally to the gravitational scalar DOF~$\phi$ through the generalized disformal metric,
a nontrivial coupling (containing second derivatives of $\phi$) given by Eq.~\eqref{eq:consistent_noninvertible} [or, equivalently, Eqs.~\eqref{eq:consistent_noninvertible_F} and~\eqref{eq:consistent_barred_metric}] is allowed.

We note that our analysis in the present paper is based on the Horndeski-frame Lagrangian, which itself makes sense even if the metric to which the matter fields are minimally coupled is not associated with invertible disformal transformations.
In this regard, it would be intriguing to take into account more general higher-order derivatives that are not covered by our transformation law~\eqref{eq:GeneralizedDisformalTransformation} (e.g., $\square\phi$) to study what types of higher derivative couplings to matter fields can survive.
It would also be interesting to investigate the consistency of matter coupling in a class of scalar-tensor theories with a nondynamical scalar field, i.e., the cuscuton~\cite{Afshordi:2006ad} or its extension~\cite{Iyonaga:2018vnu}.
These issues will be left for future work.

\acknowledgments
We thank the authors of Ref.~\cite{Naruko:2022vuh} for their helpful correspondence.
The work of TI was supported by the Rikkyo University Special Fund for Research.
The work of KT was supported by JSPS KAKENHI Grant No.~JP21J00695.
The work of TK was supported by
JSPS KAKENHI Grant No.~JP20K03936 and
MEXT-JSPS Grant-in-Aid for Transformative Research Areas (A) ``Extreme Universe'',
No.~JP21H05182 and No.~JP21H05189.

\appendix*
\section{Consistency of fermionic matter coupling}
\label{App}

In Sec.~\ref{ssec:AwayFromUnitary}, we showed that the generalized disformal transformation~\eqref{eq:GeneralizedDisformalTransformation}, with which we define the GDH theory, is restricted to be of the conventional form~\eqref{eq:ConventionalDisformalTransformation} so that a matter scalar field can be consistently coupled without introducing an extra mode, provided that the invertibility conditions~\eqref{eq:InvertibilityConditions} are satisfied.
On the other hand, our analysis was based on the Horndeski-frame Lagrangian~\eqref{eq:Horndeski-frame_Lagrangian}, which itself makes sense as a theory of gravity nonminimally coupled to matter field(s) even if the disformal transformation is noninvertible.
Interestingly, if we relax the invertibility conditions, there is a nontrivial family of generalized disformal transformations given by Eq.~\eqref{eq:consistent_barred_metric} that allows for consistent coupling of a matter scalar field, where the deviation of Eq.~\eqref{eq:consistent_barred_metric} from the conventional disformal transformation is characterized by the function~$\beta_0=\beta_0(\phi,X)$.
However, it remains unclear whether the transformation~\eqref{eq:consistent_barred_metric} with $\beta_0\ne 0$ accommodates consistent coupling of fermionic matter fields.
In this Appendix, following the discussion in Ref.~\cite{Takahashi:2022ctx}, we argue that further imposing the consistency of spinorial matter coupling leads to $\beta_0=0$, i.e., we are again left with the conventional disformal transformation.

For this purpose, one needs to study the transformation law for the tetrad under the generalized disformal transformation, since the action of fermions in curved spacetime is written in terms of the tetrad.
The authors of Ref.~\cite{Takahashi:2022ctx} developed the tetrad transformation law for the class of generalized disformal transformations defined by Eq.~\eqref{GDT_consistent_unitary}, which we repeat here for convenience:
    \begin{align}
    \bar{g}_{\mu\nu}=\tilde{F}_0g_{\mu\nu}+\tilde{F}_1\phi_\mu\phi_\nu+2\tilde{F}_2\phi_{(\mu}{\cal X}_{\nu)}+\tilde{F}_3{\cal X}_{\mu}{\cal X}_{\nu}, \qquad
    {\cal X}_{\mu}\coloneqq \left(\delta_\mu^\alpha-\frac{\phi_\mu\phi^\alpha}{X}\right)\partial_\alpha X,
    \label{GDT_consistent}
    \end{align}
where ${\cal X}_\mu$ is the derivative of $X$ projected onto a constant-$\phi$ hypersurface and $\tilde{F}_i=\tilde{F}_i(\phi,X,{\cal Z})$, with ${\cal Z}\coloneqq {\cal X}^\mu{\cal X}_\mu$.
The reason why they focused on this particular type of generalized disformal transformation is that it trivially accommodates consistent bosonic matter coupling under the unitary gauge~\cite{Takahashi:2022mew,Naruko:2022vuh}.
The analysis in Ref.~\cite{Takahashi:2022ctx} shows that $\tilde{F}_3=0$ is necessary to avoid the revival of the Ostrogradsky ghost for fermionic matter coupling.

One can recast the generalized disformal metric~\eqref{eq:consistent_barred_metric} into the form
    \begin{align}
	\bar g_{\mu\nu}=\sqrt{\alpha_0(\alpha_0+\alpha_1 X)}\bigg[&g_{\mu\nu}+\frac{\beta_0^2Y^2+2\alpha_0\beta_0Y-\alpha_0\alpha_1 X}{X\alpha_0(\alpha_0+\alpha_1 X)}\phi_\mu\phi_\nu \nonumber \\
	&+\frac{2\beta_0(\alpha_0+\beta_0Y)}{\alpha_0(\alpha_0+\alpha_1 X)}\phi_{(\mu}{\cal X}_{\nu)}+\frac{X\beta_0^2}{\alpha_0(\alpha_0+\alpha_1 X)}{\cal X}_{\mu}{\cal X}_{\nu}\bigg],
	\label{eq:consistent_noninvertible_recast}
    \end{align}
but this is not of the form~\eqref{GDT_consistent} because some of the coefficient functions depend on $Y$, which cannot be written in terms of $(\phi,X,{\cal Z})$.
Nevertheless, the discussion in Ref.~\cite{Takahashi:2022ctx} itself applies even if the coefficient functions~$\tilde{F}_i$ in Eq.~\eqref{GDT_consistent} had $Y$-dependence, as we shall see below.

In what follows, let us promote the coefficient functions~$\tilde{F}_i$ in Eq.~\eqref{GDT_consistent} as functions of $(\phi,X,Y,Z)$.
The transformation law for the tetrad~$e^a_\mu$ associated with the generalized disformal transformation can be written in the form~\cite{Takahashi:2022ctx}
    \begin{align}
    \bar{e}^a_\mu
    =\left(E_0\delta^\alpha_\mu+E_1\phi_\mu\phi^\alpha+E_2\phi_\mu {\cal X}^\alpha+E_3{\cal X}_\mu{\cal X}^\alpha\right)e^a_\alpha,
    \label{barred_tetrad}
    \end{align}
with
    \begin{align}
    E_0=\sqrt{\tilde{F}_0}, \qquad
    E_1=\frac{\sqrt{X/\bar{X}}-\sqrt{\tilde{F}_0}}{X}, \qquad
    E_2=\frac{\tilde{F}_2}{\sqrt{\tilde{F}_0+{\cal Z} \tilde{F}_3}}, \qquad
    E_3=\frac{\sqrt{\tilde{F}_0+{\cal Z} \tilde{F}_3}-\sqrt{\tilde{F}_0}}{\cal Z}.
    \end{align}
Indeed, it is straightforward to verify that $\bar{g}_{\mu\nu}=\eta_{ab}\bar{e}^a_\mu\bar{e}^b_\nu$.
Note that one could add a term~${\cal X}_\mu\phi^\alpha$ inside the parentheses in Eq.~\eqref{barred_tetrad}, but it can always be absorbed into a local Lorentz transformation~\cite{Takahashi:2022ctx}.
Having introduced the tetrad transformation law, let us consider the generalized disformal transformation of the action for a fermionic matter field represented by a free massless Dirac spinor~$\lambda$, i.e., 
    \begin{align}
    S_{\rm m}[e^a_\mu,\lambda]=\int \D^4x\,e\left(-\frac{1}{2}\lambda^\dagger {\rm i}\gamma^{\hat{0}}e^\mu_a\gamma^a\nabla_\mu\lambda+{\rm c.c.}\right), \label{fermion}
    \end{align}
where $e\coloneqq \det e^a_\mu$, ${\rm c.c.}$ denotes the complex conjugate, and $\gamma^a$ denotes the gamma matrices in the Minkowski spacetime such that $\gamma^a\gamma^b+\gamma^b\gamma^a=2\eta^{ab}\mathbbm{1}$, with $\mathbbm{1}$ being the identity matrix in the spinor indices.
Note that we put hats on local Lorentz indices ($a,b,\cdots=\{\hat{0},\hat{1},\hat{2},\hat{3}\}$).
The covariant derivative acting on the Dirac field is defined by
    \begin{align}
    \nabla_\mu\lambda
    \coloneqq \left(\mathbbm{1}\partial_\mu+\frac{1}{4}\omega_{\mu}{}^{ab}\gamma_{ab}\right)\lambda.
    \end{align}
Here, $\gamma_{ab}\coloneqq (\gamma_a\gamma_b-\gamma_b\gamma_a)/2$ and the (torsion-free) spin connection~$\omega_{\mu}{}^{ab}$ is defined by
    \begin{align}
    \omega_\mu{}^a{}_b=-e_b^\nu\left(\partial_\mu e^a_\nu-\Gamma^\alpha_{\mu\nu} e^a_\alpha\right),
    \end{align}
where $\Gamma^\lambda_{\mu\nu}$ is the Christoffel symbol associated with the metric.
We now consider the generalized disformal transformation of the spinor action~\eqref{fermion}.
Since we are only interested in the degeneracy structure of the action~\eqref{fermion}, let us focus on terms that involve time derivatives~\cite{Takahashi:2022ctx}:
    \begin{align}
    S_{\rm m}[e^a_\mu,\lambda]
    \supset \int\D^4x\sqrt{h}\left(\frac{\rm i}{2}\lambda^\dagger\dot{\lambda}-\frac{\rm i}{2}\dot{\lambda}^\dagger\lambda+\frac{\rm i}{4}\lambda^\dagger\tri_{\hat{i}}^k\tridot_{\hat{j}k}\gamma^{\hat{i}\hat{j}}\lambda\right),
    \label{spinor_kinetic}
    \end{align}
where $\tri^{\hat{i}}_k$ denotes the triad such that $h_{kl}=\delta_{\hat{i}\hat{j}}\tri^{\hat{i}}_k\tri^{\hat{j}}_l$ and $h\coloneqq \det h_{kl}=(\det \tri^{\hat{i}}_k)^2$.
Replacing the tetrad by the barred one, we obtain~\cite{Takahashi:2022ctx}
    \begin{align}
    S_{\rm m}[\bar{e}^a_\mu,\lambda]
    \supset \int\D^4x\sqrt{h}\,E_0^2(E_0+{\cal Z} E_3)\bigg[&\frac{\rm i}{2}\lambda^\dagger\dot{\lambda}-\frac{\rm i}{2}\dot{\lambda}^\dagger\lambda+\frac{\rm i}{4}\lambda^\dagger\tri_{\hat{i}}^k\tridot_{\hat{j}k}\gamma^{\hat{i}\hat{j}}\lambda \nonumber \\
    &-\frac{\rm i}{4}\frac{{\cal Z} E_3^2}{E_0(E_0+{\cal Z} E_3)}{\cal X}_m\dot{\cal X}_l\lambda^\dagger\tri_{\hat{i}}^k\tri_{\hat{j}}^l\gamma^{\hat{i}\hat{j}}\lambda\bigg]. \label{spinor_kinetic_bar}
    \end{align}
As detailed in Ref.~\cite{Takahashi:2022ctx}, under the unitary gauge, the last term inside the square brackets leads to nondegenerate higher-order derivatives in the equations of motion for the lapse function and the spinor field.
Therefore, one needs to impose $E_3=0$, i.e., $\tilde{F}_3=0$ in order not to avoid the Ostrogradsky ghost.
Note that this condition obtained under the unitary gauge should be a necessary condition for the fermionic matter coupling to be consistent in an arbitrary coordinate system.
Since the transformation~\eqref{eq:consistent_noninvertible_recast} has $\tilde{F}_3\propto \beta_0^2$, the condition~$\tilde{F}_3=0$ requires $\beta_0=0$.

\bibliography{refs}
\bibliographystyle{JHEP}
\end{document}